\documentclass[twoside,twocolumn,9pt]{article}
\usepackage{extsizes}
\usepackage[super,sort&compress,comma]{natbib} 
\usepackage[version=3]{mhchem}
\usepackage[left=1.5cm, right=1.5cm, top=1.785cm, bottom=2.0cm]{geometry}
\usepackage{balance}
\usepackage{times,mathptmx}
\usepackage{sectsty}
\usepackage{graphicx} 
\usepackage{lastpage}
\usepackage[format=plain,justification=justified,singlelinecheck=false,font={stretch=1.125,small,sf},labelfont=bf,labelsep=space]{caption}
\usepackage{float}
\usepackage{fancyhdr}
\usepackage{fnpos}
\usepackage[english]{babel}
\addto{\captionsenglish}{%
  
}
\usepackage{array}
\usepackage{droidsans}
\usepackage{charter}
\usepackage[T1]{fontenc}
\usepackage[usenames,dvipsnames]{xcolor}
\usepackage{setspace}
\usepackage[compact]{titlesec}
\usepackage{hyperref}
\usepackage{amsmath}
\usepackage{amsfonts}
\usepackage{amssymb}
\usepackage{bm}

\newcommand*{\citen}[1]{%
  \begingroup
    \romannumeral-`\x % remove space at the beginning of \setcitestyle
    \setcitestyle{numbers}%
    \cite{#1}%
  \endgroup
}

\definecolor{cream}{RGB}{222,217,201}

\begin{document}

\pagestyle{plain}
\thispagestyle{plain}

%%%PAGE SETUP - Please do not change any commands within this section%%%
\makeFNbottom
\makeatletter
\renewcommand\LARGE{\@setfontsize\LARGE{15pt}{17}}
\renewcommand\Large{\@setfontsize\Large{12pt}{14}}
\renewcommand\large{\@setfontsize\large{10pt}{12}}
\renewcommand\footnotesize{\@setfontsize\footnotesize{7pt}{10}}
\makeatother

\renewcommand{\thefootnote}{\fnsymbol{footnote}}
\renewcommand\footnoterule{\vspace*{1pt}% 
\color{cream}\hrule width 3.5in height 0.4pt \color{black}\vspace*{5pt}} 
\setcounter{secnumdepth}{5}

\makeatletter 
\renewcommand\@biblabel[1]{#1}
\renewcommand\@makefntext[1]% 
{\noindent\makebox[0pt][r]{\@thefnmark\,}#1}
\makeatother 
\renewcommand{\figurename}{\small{Fig.}~}
\sectionfont{\sffamily\Large}
\subsectionfont{\normalsize}
\subsubsectionfont{\bf}
\setstretch{1.125} %In particular, please do not alter this line.
\setlength{\skip\footins}{0.8cm}
\setlength{\footnotesep}{0.25cm}
\setlength{\jot}{10pt}
\titlespacing*{\section}{0pt}{4pt}{4pt}
\titlespacing*{\subsection}{0pt}{15pt}{1pt}
%%%END OF PAGE SETUP%%%

%%%FOOTER%%%
\fancyfoot{}
\fancyfoot[RO]{\footnotesize{\sffamily{1--\pageref{LastPage} ~\textbar  \hspace{2pt}\thepage}}}
\fancyfoot[LE]{\footnotesize{\sffamily{\thepage~\textbar\hspace{3.45cm} 1--\pageref{LastPage}}}}
\fancyhead{}
\renewcommand{\headrulewidth}{0pt} 
\renewcommand{\footrulewidth}{0pt}
\setlength{\arrayrulewidth}{1pt}
\setlength{\columnsep}{6.5mm}
\setlength\bibsep{1pt}
%%%END OF FOOTER%%%

%%%FIGURE SETUP - please do not change any commands within this section%%%
\makeatletter 
\newlength{\figrulesep} 
\setlength{\figrulesep}{0.5\textfloatsep} 

\newcommand{\topfigrule}{\vspace*{-1pt}% 
\noindent{\color{cream}\rule[-\figrulesep]{\columnwidth}{1.5pt}} }

\newcommand{\botfigrule}{\vspace*{-2pt}% 
\noindent{\color{cream}\rule[\figrulesep]{\columnwidth}{1.5pt}} }

\newcommand{\dblfigrule}{\vspace*{-1pt}% 
\noindent{\color{cream}\rule[-\figrulesep]{\textwidth}{1.5pt}} }

\makeatother
%%%END OF FIGURE SETUP%%%

%%%TITLE, AUTHORS AND ABSTRACT%%%
\twocolumn[
  \begin{@twocolumnfalse}
  \LARGE{\textbf{Clustering and phase separation of circle swimmers dispersed in a monolayer$^\dag$}} \\
  \\
  \large{Guo-Jun Liao$^{\ast}$\textit{$^{a}$} and Sabine H. L. Klapp\textit{$^{a}$}} \\
  \\
  \normalsize{We perform Brownian dynamics simulations in two dimensions to study the collective behavior of circle swimmers, which are driven by both, an (effective) translational and rotational self-propulsion, and interact via steric repulsion. We find that active rotation generally opposes motility-induced clustering and phase separation, as demonstrated by a narrowing of the coexistence region upon increase of the propulsion angular velocity. Moreover, although the particles are intrinsically assigned to rotate counterclockwise, a novel state of clockwise vortices emerges at an optimal value of the effective propulsion torque. We propose a simple gear-like model to capture the underlying mechanism of the clockwise vortices.} \\
 \end{@twocolumnfalse} \vspace{0.6cm}

  ]
%%%END OF TITLE, AUTHORS AND ABSTRACT%%%

%%%FONT SETUP - please do not change any commands within this section
\renewcommand*\rmdefault{bch}\normalfont\upshape
\rmfamily
\section*{}
\vspace{-1cm}

%%%FOOTNOTES%%%

\footnotetext{\textit{$^{a}$~
Institut f\"ur Theoretische Physik, Technische Universit\"at Berlin, Hardenbergstr. 36, D-10623 Berlin, Germany. E-mail: guo-jun.liao@campus.tu-berlin.de}}

%Please use \dag to cite the ESI in the main text of the article.
\footnotetext{\dag~Due to the limit of file size, Electronic Supplementary Information (ESI) except Movie3.avi is only available via request.}

%%%END OF FOOTNOTES%%%

%%%MAIN TEXT%%%%
\section{\label{sec:intro} Introduction}
Self-propelled (active) particles exhibit a wealth of intriguing collective states, including clustering, \cite{Theurkauff2012, Cates2013, Bialke2015} swarming, \cite{Vicsek1995, Buhl2006, Zhang2009, Copeland2009} swirling, \cite{Dombrowski2004, Kudrolli2008} laning, \cite{Kogler2015, Wachtler2016} and mesoscale turbulence. \cite{Wensink2012, Heidenreich2016, Reinken2018}
Examples occur over a wide range of length and time scales, including pedestrians, \cite{Helbing1995} bacteria, \cite{Maeda1976} and self-propelled bimetallic nanorods, \cite{Paxton2004} or Janus particles. \cite{Gangwal2008} 
Depending on the type of interactions among the active particles, their motility-induced macroscopic structures can be well described by surprisingly simple models. 
To name two prominent models, the Vicsek model captures active particles favoring parallel alignment of propulsion directions, \cite{Vicsek1995} and the conventional Active Brownian Particle (ABP) model is adequate for self-propelled particles interacting via isotropic (spherical or disk-like) repulsion due to excluded volume. \cite{Buttinoni2013} In the present paper we consider a variant of the ABP model.
%

% New Paragraph
% 
In a suspension of conventional ABPs, even though the interactions are purely repulsive, the system can undergo a transition from a homogenous disordered state to a state characterized by coexistence of clusters and freely moving swimmers, as the particle motility increases. \mbox{\cite{Buttinoni2013, Stenhammar2013, Cates2015}}
This transition resembles the liquid-gas phase separation of equilibrium fluids with attractive interactions, if one views the particle motility in the active system as an analog to the attractive coupling in the passive fluid.
The clustering phenomenon is therefore often referred to as ``motility-induced phase separation.'' \cite{Cates2015}
%

% New Paragraph
%
Many studies of self-propelled particles consider the case in which the effective propulsion force of a single particle coincides with its center of mass (due to its shape symmetry), thereby causing the particle to move along a straight line perturbed by thermal fluctuations. \cite{Yang2010, Wensink2012a, Buttinoni2012a, Redner2013a} 
However, for an asymmetric active particle, the net effective propulsion force does not coincide with its center of mass, thus additionally introducing a effective propulsion torque. 
The interplay of effective propulsion force and torque induces a ``circular motion'' of the asymmetric particle in the absence of thermal fluctuations. Therefore, such active particles are referred to as ``circle swimmers.'' \cite{VanTeeffelen2008}
Real-world examples of circle swimmers include \textit{E.coli} which swims clockwise upon in contact with interfaces, \cite{Lauga2006} as well as FtsZ proteins, which exhibit clockwise treadmilling on membranes. \cite{Loose2014} 
Circle swimmers can also be artificially prepared, such as L-shaped particles \cite{Kummel2013} and Janus particles with asymmetric coating on the surface. \cite{Mano2017, Campbell2017}

% New Paragraph 
%
There are several studies in the literature which have addressed the dynamics of circle swimmers.
For instance, the motion of a single circle swimmer has been investigated in detail both theoretically, \cite{VanTeeffelen2008, Ao2015, Sevilla2016, Kurzthaler2017, Jahanshahi2017} and in experiments. \cite{Kummel2013} 
It has also been demonstrated that non-interacting chiral microswimmers can be sorted in presence of chiral static obstacles. \cite{Mijalkov2013}
Examples of interacting circle swimmers include purely repulsive, athermal active disks. \cite{Lei2018}
More specifically, circle swimmers may align with their neighbors via anisotropic swimmer shapes, \cite{Kaiser2013, Denk2016} local gradient of chemical concentration, \cite{Liebchen2016} or imposed alignment interactions. \cite{Liebchen2017, Levis2018} 
However, the overall collective behavior of chiral active Brownian disks without any alignment mechanism is not yet fully understood. 
This concerns, in particular, the impact of active rotation on the phase separation.
%

%% New Paragraph
%
Motivated by this lack, we study in the present work the conventional ABP model with an additional active rotation term. 
Different from the athermal case, \cite{Lei2018} the propulsion direction of each Brownian swimmer is subject not only to an intrinsic effective propulsion torque but also to thermal fluctuations.
As a result, an individual swimmer moves along a circular path, perturbed by thermal noise.
Based on this model, we explore via Brownian dynamics simulations the occurrence of cluster formation and phase separation.
%

%% New Paragraph
%
As we will demonstrate, the competition of active rotation and thermal noise indeed dominates the collective dynamics at high densities.
Giant clusters and phase separation only appear when the angular speed is much smaller than the rotational diffusion. 
At larger angular speeds, we observe a drastic decrease of the size of the largest cluster, accompanied by the emergence of clockwise vortices. 
This behavior of clockwise rotation is somewhat counterintuitive, since freely moving particles rotate, by default, counterclockwise. 
We propose a simple argument to capture the underlying mechanism of this intriguing behavior.
%

%% Organization of this paper
The rest of this paper is organized as follows. 
In Sec.~\ref{sec:model} we present our model of circle swimmers and the details of Brownian dynamics simulations. 
Based on the simulation results we discuss in Sec.~\ref{sec:resultA} and~\ref{sec:resultB} the influence of the strength of active rotation on the motility-induced clustering and phase separation.
At specific values of active rotation, clockwise vortices appear, which we discuss in Sec.~\ref{sec:resultC}.
Finally, we summarize our conclusions in Sec.~\ref{sec:conclusions}.

\section{\label{sec:model} Model and simulation techniques}

We perform Brownian dynamics simulations with $N$ circle swimmers in two dimensions (2D) in the $xy$-plane. 
Neglecting hydrodynamic interactions, the overdamped motion of the $i$th swimmer is computed by solving the coupled Langevin equations \cite{VanTeeffelen2008} for its center-of-mass position $\boldsymbol{r}_i$ and orientation $\widehat{\boldsymbol{e}}_i = \left( \text{cos}\psi_i\text{, } \text{sin}\psi_i \right)^{T}$
\begin{align} 
  \dot{\boldsymbol{r}}_i & = 
    \beta \mathbb{D}
      \Big[
        F_{0}\widehat{\boldsymbol{e}}_{i} 
        - \nabla_{\boldsymbol{r}_i}U 
        + \boldsymbol{\xi}_{i}\left(t\right) 
      \Big], \label{eqn:coupled_Langevin_trans}\\
  \dot{\psi}_i & = 
    \beta D_r
      \Big[
        M_{0} 
        - \partial_{\psi_i} U 
        + \Gamma_{i}\left(t\right)
      \Big], \label{eqn:coupled_Langevin_rot}
\end{align}
where the dots denote time derivatives and $\beta^{-1} = k_{B}T$ is the thermal energy (with $k_{B}$ being Boltzmann's constant and $T$ being the temperature). 
Each swimmer is modeled as a disk; it is thus isotropic in shape. 
We therefore set the translational diffusion tensor $\mathbb{D} = D_{t} \mathbb{I}$, where $D_t$ is the unit of translational diffusion constant and $\mathbb{I}$ denotes the $2 \times 2$ identity matrix.
Further, $D_{r}$ represents the rotational diffusion constant, and $F_{0}\widehat{\boldsymbol{e}}_i$ and $M_{0}$ are the effective propulsion force and torque which drive the active motion of the $i$th swimmer with a preferred direction of rotation.
For $M_{0} = 0$ the model equations \eqref{eqn:coupled_Langevin_rot} reduce to those of conventional ABPs.
In the remainder of the paper, we describe the impact of the effective propulsion force and torque via the motility $v_0 = \beta D_t F_0$ and the angular speed $\omega_0 = \beta D_r M_0$.
The thermal fluctuations due to the collisions of solvent molecules are represented by the random force 
$\boldsymbol{\xi}_i(t)$ and the random torque $\Gamma_i(t)$, respectively, which are zero mean Gaussian white noises with temporal correlations 
$\langle\xi_{i, \mu}(t)\xi_{j, \nu}(t')\rangle = 2\delta_{ij} \delta_{\mu \nu} \delta(t-t')/(D_t\beta^{2})$ and $\langle\Gamma_i(t)\Gamma_j(t')\rangle = 2\delta_{ij}\delta(t-t')/(D_r\beta^{2})$. 
Here, $\xi_{i, \mu}(t)$ is the $\mu$ ($x$ or $y$) component of $\boldsymbol{\xi}_i(t)$ for the $i$th particle. 
Angle brackets denote ensemble average. 
%

%% New Paragraph
%
In the present study, the particle interaction $U$ appearing in eqn \eqref{eqn:coupled_Langevin_trans} and \eqref{eqn:coupled_Langevin_rot} describes (only) steric repulsion. Specifically, we employ the Weeks-Chandler-Andersen (WCA) potential \cite{Weeks1971} between particles $i \neq j$. The functional form is given by
\begin{equation} \label{eqn:wca}
U_{WCA}(r_{ij})= 
  \begin{cases}
    4\epsilon \left[
      \left(\dfrac{\sigma}{r_{ij}}\right)^{12} - 
      \left(\dfrac{\sigma}{r_{ij}}\right)^{6} + 
      \dfrac{1}{4}
    \right]
    \text{,} &\text{if $r_{ij} < r_{c}$,}\\
    0\text{,} &\text{else,}
  \end{cases}
\end{equation}
with the particle distance $r_{ij} = \vert \boldsymbol{r}_{ij} \vert = \vert \boldsymbol{r}_{j} - \boldsymbol{r}_{i} \vert$. We choose the repulsive strength $\epsilon^{*} = \beta \epsilon = 100$ and set the unit of length to be $\sigma$. The potential is truncated at a cut-off (c) distance $r_{c} =2^{1/6}\sigma$. The potential $U_{WCA}$ and the force $-\nabla U_{WCA}$ are continuous at the truncation point, and their values are zero as $r_{ij} \geq r_{c}$. 
%

% New Paragraph
%
The translational and rotational diffusion constants are related via $D_r = 3D_t/\sigma_h^2$, as found by solving the Navier-Stokes-equation for a hard-spherical particle of diameter $\sigma_h$ in the low Reynolds number regime. \cite{Landau1987} 
Following the treatment proposed by Barker and Henderson, \cite{Barker1967, Ilg2009} we can define an effective (eff) hard sphere diameter via $\sigma_{eff} = \int_0^{\infty}\left(1-\text{exp}\left[-\beta U_{WCA}\left(r\right)\right]\right)\text{d}r$.
At the repulsion strength $\epsilon^* = 100$ considered here, $\sigma_{eff} \approx 2^{1/6}\sigma$.
Choosing $\sigma_{h} = \sigma_{eff} = 2^{1/6}\sigma$ we thus obtain $D_r = 3 \times 2^{-1/3} D_t/\sigma^2$.
%

%% New Paragraph
%
Finally, it is instructive to briefly recall the behavior of a single particle governed by eqn \eqref{eqn:coupled_Langevin_trans} and \eqref{eqn:coupled_Langevin_rot} with $U = 0$.
This case has already been analyzed in ref. \citen{VanTeeffelen2008}.
In the absence of thermal fluctuations (\textit{i.e.}, $\xi_{\mu}\left(t\right) = \Gamma\left(t\right) = 0$), the particle moves on a circle with radius $R_0 = v_0 / \omega_0$.
With thermal fluctuations, the circle transforms into a logarithmic spiral. \cite{VanTeeffelen2008}
%

%  New Paragraph
%
All simulations are performed with at least $4 \times 10^6$ time steps. 
The time difference between each time step is $\Delta t = 1\times 10^{-5} \tau$ at the largest with the time unit $\tau = \sigma^2/D_{t}$. 
In order to obtain steady-state results, the system properties are measured after the first half of a simulation is performed, i.e. after at least $2 \times 10^6$ time steps. 
In the second half of a simulation, we take ``snapshots'' of particle positions and orientations every 1000 time steps.
Averages are then calculated on the basis of these snapshots.
Each data point shown in Sec. \ref{sec:result} is carried out for a single simulation.
We use $N = 5000$ circle swimmers in a quadratic box ($L \times L$) with periodic boundary conditions.
The mean area fraction is defined as $\Phi = N \pi \sigma^2 / (4L^2)$. 
The simulation results are presented in dimensionless units. 
The dimensionless propulsion speed is given by $v_0^* = v_0 \sigma / D_t$. 
To characterize the strength of active rotation compared to that of thermal fluctuations, we define the dimensionless propulsion angular speed by $\omega_0^* = \omega_0/D_r$. 
As $0 < \omega_0^* < 1$, thermal fluctuations dominate, and at $\omega_0^* > 1$ vice versa.
We choose a positive angular speed $\omega_{0}^{*} > 0$ such that a single particle rotates counterclockwise. 
\begin{figure}[!htbp]
  \centering
  \includegraphics[width=0.9\linewidth]{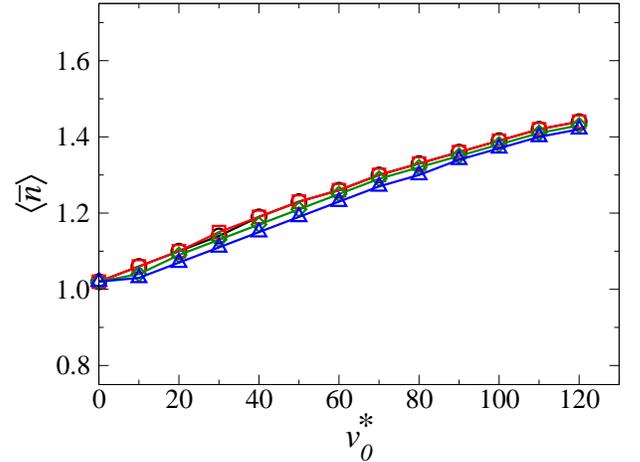}
  \caption[small clusters of circle swimmers]{(Color online) Mean cluster size $\langle \bar{n} \rangle$ as a function of swimmer motility $v_{0}^{*}$ for angular speed $\omega_0^* = 0$ (black circles), $1$ (red squares), $6$ (green diamonds), and $10$ (blue triangles) at $\Phi = 0.2$. The solid lines are guides to the eye.}
  \label{fig:meanClust}
\end{figure}

\section{\label{sec:result} Numerical Results}
\subsection{\label{sec:resultA} Cluster formation}
%% New Paragraph 
%
At sufficiently high density and particle motility, conventional ABPs tend to form clusters and eventually phase separate into dilute and dense (cluster-dominated) regions \cite{Theurkauff2012, Cates2013, Bialke2015, Buttinoni2013, Stenhammar2013, Cates2015}.
This phenomenon can be understood by the following picture: Considering two identical conventional ABPs with opposite orientation bumping into each other head-to-head, they mutually annihilate the translational propulsion and stop moving for a short period of time.
This pair of ABPs then becomes a temporary obstacle to the neighboring swimmers, which collide with the obstacle and are thus slowed down.
As the effective propulsion force increases, the swimmer motility becomes higher.
This leads to more collisions per unit time, and thus, a more significant slowing-down effect, thereby causing swimmers to form clusters.
%

%% New Paragraph
%
A cluster is considered to be stable once the rate of the number of particles joining in and escaping from this cluster are balanced. 
An active particle escapes from the cluster surface once its orientation changes from pointing inward into the cluster to outward. \cite{Redner2013, Redner2013a}
For a conventional ABP, the rotational diffusion is the only escaping mechanism to alter its direction of effective propulsion force.
In contrast, a circle swimmer varies its orientation via not only the rotational diffusion but also the active rotation.
Therefore, it is interesting to see how the active rotation influences the motility-induced cluster formation.
%

% New Paragraph
%
In our study we determine clusters via a distance criterion: \mbox{\cite{Buttinoni2013, Bialke2013}} 
The \textit{i}th particle is regarded as being in contact with the \textit{j}th particle if $r_{ij} = \vert \boldsymbol{r}_{ij} \vert = \vert \boldsymbol{r}_{j} - \boldsymbol{r}_{i}\vert \leq \sigma_{eff}$ (see the text below eqn~\eqref{eqn:wca}). 
A cluster is then a set of particles that are in contact with each other, and its size $n$ is defined as number of particles therein.
Further technical details on the cluster analysis, particularly concerning the calculation of averages, are given in Appendix \ref{app:steady_state}.

%% New Paragraph
%
To characterize the cluster formation in the low density regime (e.g. $\Phi = 0.2$), we plot in Fig.~\ref{fig:meanClust} the mean cluster size $\langle \bar{n} \rangle$ as a function of swimmer motility $v_0^*$ for various values of angular speed $\omega_0^*$.
The limiting case is $\omega_0^* = 0$, where our model reduces to conventional ABPs. For that case, we find the mean cluster size $\langle \bar{n} \rangle$ to increase linearly with $v_0^*$. No apparent giant cluster is observed.
Such linear relationship is also observed in ref. \citen{Theurkauff2012} and \citen{Buttinoni2013}.
However, the mean cluster size $\langle \bar{n} \rangle$ does not increase as rapidly as in the experiments described in ref. \citen{Theurkauff2012} and \citen{Buttinoni2013}.
Indeed, in the experiments, the dynamical clustering is presumably strongly influenced by phoretic/chemical interactions among the particles, \cite{Theurkauff2012} which are not included in our ABP-like model.
The mean cluster size $\langle \bar{n} \rangle$ keeps approximately unchanged for different $\omega_0^*$, even when the active rotation dominates rotational diffusion ($\omega_0^* = 10$).
\begin{figure}[!htbp]
  \centering
  \includegraphics[width=0.85\linewidth]{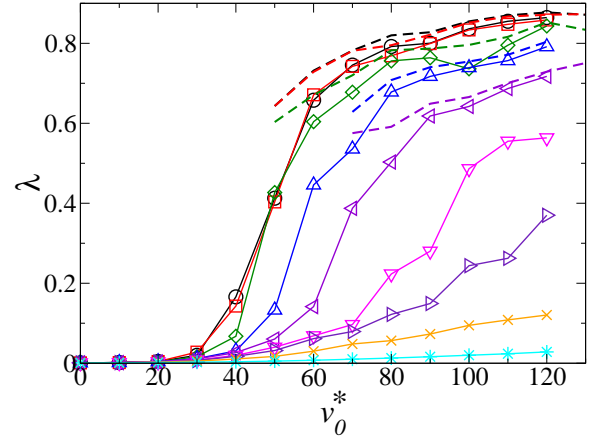}
  \caption[Clustering of swimmers]{(Color online) Fraction of the largest cluster $\lambda$ as a function of motility $v_0^*$ for angular speed $\omega_0^* = 0$ (red squares), $0.1$ (green diamonds), $0.6$ (blue triangles up ), $0.8$ (violet triangles left), $1.0$ (magenta triangles down), $1.2$ (indigo triangles right), $2.0$ (orange crosses), and $10$ (cyan stars) at $\Phi = 0.4$. The data of $\omega_0^* = 0$ (red squares) are quantitatively in good agreement with ref. \citen{Bialke2013} (black circles). As will be discussed later in Sec. \ref{sec:resultB}, dashed lines are plotted according to eqn~\eqref{eqn:lever_rule}.}
  \label{fig:clustr_w0}
\end{figure}
%

% New Paragraph
%
At higher densities such as $\Phi = 0.4$, circle swimmers with sufficiently large motility undergo a transition from many small clusters into densely packed giant clusters coexisting with freely moving swimmers (if $\omega_0^*$ is not too large). 
In this regime of densities, analyzing the average size of the largest cluster (lcl), $\langle n_{\text{lcl}} \rangle$, becomes more physically meaningful than measuring the mean cluster size $\langle \bar{n} \rangle$. 
In Fig.~\ref{fig:clustr_w0} we plot the fraction of the largest cluster $\lambda = \langle n_{\text{lcl}} \rangle / N$ as a function of motility $v_0^*$ for various angular speeds $\omega_0^*$.
The data for $\omega_0^* = 0$ are in good agreement with the reference data for ABPs in ref. \citen{Bialke2013}.
Upon increase of $\omega_0^*$, we see that the sharp increase of $\lambda$ at $v_0^* \approx 40$ progressively weakens.
In fact, the values of $\lambda$ significantly decrease as $\omega_0^* \approx 1$, where the active rotation becomes of the same order of magnitude as the rotational diffusion. 
For $\omega_0^* \geq 1$, the active rotation is the dominant mechanism for altering the particle orientation, such that particles ``escape'' from the cluster more easily than merely relying on rotational thermal fluctuations. 
Therefore, the clusters become unstable and the fraction of the largest cluster decreases.
%

\begin{figure}[!htbp]
  \centering
  \includegraphics[width=0.9\linewidth]{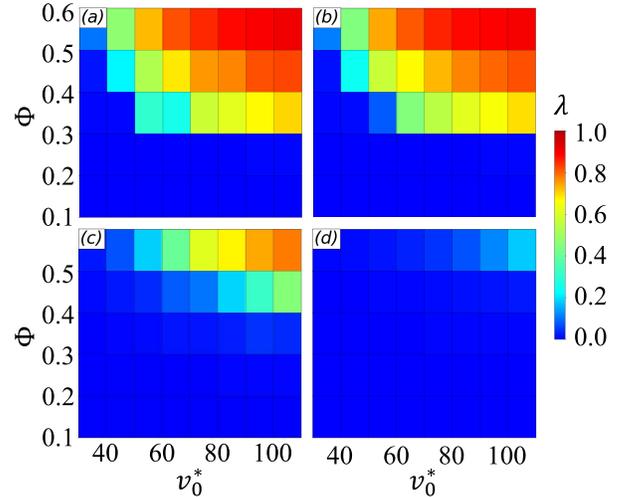}
  \caption[Phase diagrams]{(Color online) Color map of the fraction of the largest cluster in the ($\Phi$, $v_0^*$) plane for angular speed $\omega_0^{*} = 0$ (a), $0.1$ (b), $1$ (c), and $10$ (d).}
  \label{fig:4PhaseDiagram_new}
\end{figure}

%% New Paragraph
%
To illustrate the impact of the density, we plot in Fig.~\ref{fig:4PhaseDiagram_new} color maps of $\lambda$ in the ($\Phi$, $v_0^*$) plane for four different values of $\omega_0^*$. 
For $\omega_0^* = 0$, giant clusters form in the range $\Phi \gtrsim 0.3$ and $v_0^* \gtrsim 50$, as shown by the large values of $\lambda$ ($\lambda \gtrsim 0.5$) in the corresponding regions in Fig.~\ref{fig:4PhaseDiagram_new}(a).
The behavior at $\omega_0^* = 0.1$ is very similar (see Fig.~\ref{fig:4PhaseDiagram_new}(b)).
In contrast, Fig.~\ref{fig:4PhaseDiagram_new}(c) for $\omega_0^* = 1.0$ shows that the region of $\lambda \geq 0.5$ in the color map is substantially smaller.
A further increase of $\omega_0^*$ to $10$ yields a vanishing of giant clusters in the scanned parameters range of $\Phi$ and $v_0^*$, as depicted in Fig.~\ref{fig:4PhaseDiagram_new}(d).

\subsection{\label{sec:resultB} Phase separation}

Motility-induced phase separation is characterized by the presence of freely moving swimmers coexisting with dense clusters. 
Since the parameter $\lambda$ alone does not describe this type of coexistence, we compute a histogram of position-resolved, time-averaged local area fractions.
To this end we use a Voronoi tessellation. \cite{Blaschke2016}
Coexisting states are characterized by a double-peak structure of the histogram.
%

%% New Paragraph
%
In our implementation of the Voronoi tessellation, we take into account eight closest images of the central simulation box in addition to the main one, such that the space in the central box is properly partitioned.
The particle-resolved local area fraction of the $i$th particle is defined as $\phi_i = \pi \sigma^2 / \left( 4 A_i \right)$, where $A_i$ is the area 
of the $i$th Voronoi cell.
Based on $\phi_i$, we set up a grid to obtain the position-resolved local area fraction $\phi\left(x,y\right)$.
The mesh size is given by $\Delta L = L / \text{floor}(L) \approx 1 \sigma$, \footnote[3]{The floor function $\text{floor}(x)$ map a real number $x$ to the largest integer less than or equal to $x$.} which is large enough to preserve the particle-resolved information.
For each grid point $\left(x,y\right)$ located inside the $i$th Voronoi cell, we assign $\phi\left(x,y\right) \equiv \phi_i$.
We take a short-time average of $\bar{\phi}\left(x,y\right)$ over the time interval $\Delta \tau = 0.5 \tau$ to filter out the transient small clusters in dilute region. \cite{Blaschke2016}
Within this time interval, giant clusters keep the same shape, but the transient small clusters in the dilute region vanish.
As a result, the interface between the dense and dilute region is correctly recognized.
This point is crucial for determining coexistence densities, as we will state later.
\begin{figure}[!htbp]
  \centering
  \includegraphics[width=0.9\linewidth]{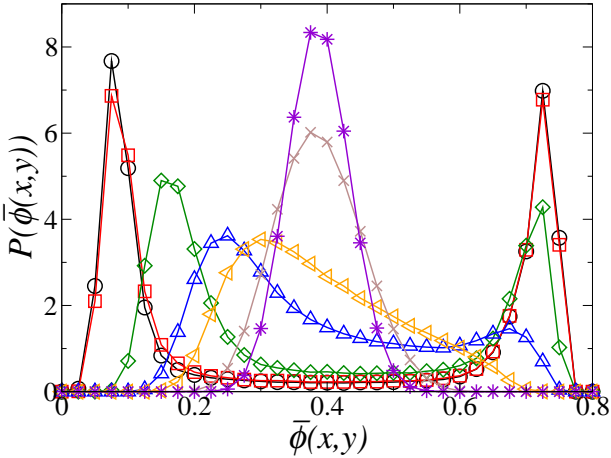}
  \caption{(Color online) Histogram of the position-resolved local area fraction $P\left(\bar{\phi}\left(x,y\right)\right)$ for angular speed $\omega_0^* = 0$ (black circles), $0.1$ (red squares), $0.8$ (green diamonds), $1.2$ (blue triangles up), $2$ (orange triangles left), $6$ (brown crosses), and $10$ (violet stars) at mean area fraction $\Phi = 0.4$ and swimmer motility $v_0^* = 120$. The solid lines are drawn as a guide to the eye.}
  \label{fig:prob_distri_pos_res}
\end{figure}
\begin{figure}[!htbp]
  \centering
  \includegraphics[width=0.9\linewidth]{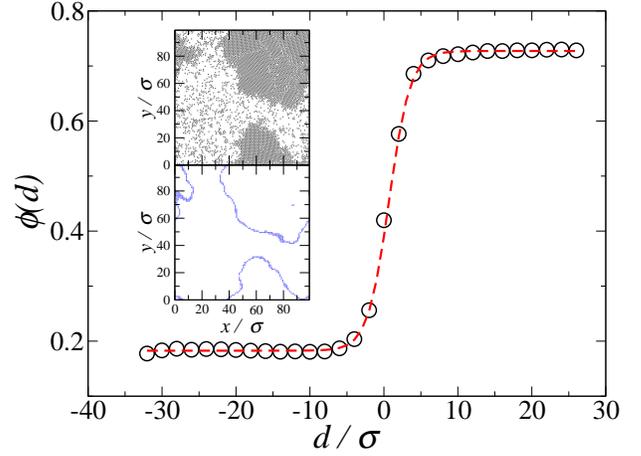}
  \caption{(Color online) The local area fraction $\phi\left(d\right)$ at the position $d$ relative to the interface from Brownian dynamics simulations (black circles) at $\Phi = 0.4$, $v_0^* = 120$, and $\omega_0^* = 0.8$. The red dashed line is the curve fitted by eqn~\eqref{eqn:fitting} with $\phi_{gas} = 0.7273$, $\phi_{gas} = 0.1827$, $d_{0} = 0.6951 \sigma$, and $w = 2.867 \sigma$. Inset: a snapshot which displays the particles as black dots (upper part) and a snapshot which shows the corresponding time-averaged interfacial grid points as blue dots with $\Delta \tau = 0.5 \tau$ (lower part).}
  \label{fig:interface}
\end{figure}
%

% New Paragraph
%
An example for the probability distribution of the local area fraction $P\left(\bar{\phi}\left(x,y\right)\right)$ is given in Fig.~\ref{fig:prob_distri_pos_res}.
It is seen that a double-peak structure and, thus, state coexistence occurs in the range $0 \lesssim \omega_0^* \lesssim 1.2$.
For $\omega_0^* \gtrsim 1.2$, the histogram reveals only one peak, indicating the absence of global phase separation.
However, the spatial structure is still inhomogeneous, as an inspection of simulation snapshots reveals.
%

% New Paragraph
%
In order to determine the densities corresponding to coexisting states, we need to measure the local area fractions far away from both sides of the interface. 
The details of determining the interface as well as the interfacial grid points are given in Appendix~\ref{app:interface}.
Figure~\ref{fig:interface} provides an example of the local area fraction as a function of the position $d$ relative to the interface, where $d < 0$ denotes the dilute region and $d > 0$ represents the dense region.
The upper inset of Fig.~\ref{fig:interface} presents an exemplary simulation snapshot of state coexistence.
From the corresponding interfacial grid points shown in the bottom inset of Fig.~\ref{fig:interface} , we see that the interface is correctly identified.
Inspired by ref. \citen{Blaschke2016} and \citen{Bialke2015a}, we fit our data with the Cahn-Hilliard \cite{Cahn1958} ansatz:
\begin{equation}
\label{eqn:fitting}
\phi\left(d\right) = 
  \frac{\phi_{den} + \phi_{gas}}{2} + 
  \frac{\phi_{den} - \phi_{gas}}{2} 
    \text{tanh} 
      \left( 
        \frac{d - d_0}{w} 
      \right)
    \text{.}
\end{equation}
This equation then yields the coexisting area fractions $\phi_{gas}$ and $\phi_{den}$. 
%

\begin{figure}[!htbp]
  \centering
  \includegraphics[width=0.9\linewidth]{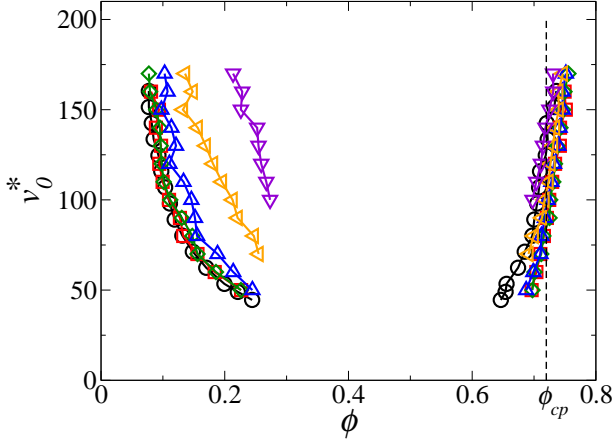}
  \caption[State Coexistence]{(Color online) Area fractions corresponding to coexisting states in the ($v_0^*$, $\phi$) plane for angular speed $\omega_0^* = 0$ (red squares), $0.1$ (green diamonds), $0.4$ (blue triangles up ), $0.8$ (orange triangles left), and $1.2$ (purple triangles down) at mean area fraction $\Phi = 0.4$. The reference data (black circles) are taken from ref. \citen{Bialke2015a} with $\omega_0^* = 0$, $\Phi = 0.3969$ ($\Phi_{\text{Bialk\'{e}}} = \pi a^2/\left(4 L^2 \right) = 0.5$ and $a = 2^{1/6}\sigma$), and $N = 10000$. The black dashed line marks the area fraction related to close-packing, $\phi_{cp} = \pi \sigma^2 \sigma_{eff}^{-2}/\left(2\sqrt{3}\right) \approx 0.72$.}
  \label{fig:phase_coexist}
\end{figure}

%% New Paragraph
%
To understand how active rotation influences the coexisting densities, we plot in Fig.~\ref{fig:phase_coexist} the resulting binodal curves in the plane of motility $v_0^*$ versus area fraction $\phi$ with various angular speeds $\omega_0^*$.
In the limiting case $\omega_0^* = 0$, our result for the gas-like, low density branch $\phi_{gas}$ agrees well with data given in ref. \citen{Bialke2015a}. 
The small deviation between our results and the reference data \cite{Bialke2015a} regarding the high density branch $\phi_{den}$ may be attributed to the different method employed to compute the local area fraction. 
%

%% New Paragraph
%
At high densities, the swimmers form a densely packed structure with the distance between two neighboring swimmers being approximately equal to the effective hard sphere diameter $\sigma_{eff}$.
Therefore, the value of the local density is close to its maximum, that is, the close-packing fraction $\phi_{cp} = \pi \sigma^{2} \sigma_{eff}^{-2}/\left(2\sqrt{3}\right) \approx 0.72$.
%

%% New Paragraph
%
At small angular speeds $\left( \omega_0^* \lesssim 0.1 \right)$, the branch corresponding to the gas-like phase $\phi_{gas}$ remain unchanged.
However, for $\omega_0^* \gtrsim 0.4$ the $\phi_{gas}$-curves are significantly shifted toward higher densities.
This shift may be attributed to the increasing ability of the circle swimmers in the dense region to alter their orientations via active rotation and subsequently return to the dilute region.
In contrast, the high-density branch slightly moves toward lower densities (upon increase of $\omega_0^*$). 
Altogether, the difference $\phi_{den} - \phi_{gas}$ decreases upon increase of angular speed (in the range $\omega_0^* \leq 1.2$), suggesting that active rotation generally opposes motility-induced phase separation.
Once $\omega_0^* > 1.2$, the coexistence of freely moving particles and stable clusters disappears, at least for the values of $v_0^*$ considered here.
%

%% New Paragraph
%
To make a connection between the state coexistence and the fraction of the largest cluster $\lambda$ considered in Fig.~\ref{fig:clustr_w0}, we assume that the mean area fractions in the dilute and dense region are $\phi_{gas}$ and $\phi_{den}$, respectively.
The areas of the dilute and dense regions are thus given by
\begin{align}
A_{gas} & = \frac{\left( 1 - m \right) N \pi\sigma^2}{4 \phi_{gas}}\text{ and} \\
A_{den} & = \frac{mN \pi\sigma^2}{4 \phi_{den}}\text{,}
\end{align}
where $m$ denotes the fraction of particles in the dense region. 
Since $L^2 = N \pi\sigma^2/\left(4\Phi\right) = A_{gas} + A_{den}$, $m$ can be expressed as \cite{Speck2015}
\begin{equation} \label{eqn:lever_rule}
m=\frac{1/\phi_{gas} - 1/\Phi}{1/\phi_{gas} - 1/\phi_{den}}.
\end{equation}
As motility-induced phase separation occurs, most particles in the dense region belong to the largest cluster, i.e. $m \approx \lambda$.
This is confirmed by the dashed lines plotted in Fig.~\ref{fig:clustr_w0}, which have been calculated from eqn~\eqref{eqn:lever_rule} and obviously agree well with the data obtained from the cluster analysis.
The slight overestimation predicted by eqn~\eqref{eqn:lever_rule} may result from the fact that not every particle in the dense region resides in the largest cluster. Furthermore, eqn~\eqref{eqn:lever_rule} by construction neglects interfacial regions.
%

%% New Paragraph
%
To conclude, Fig.~\ref{fig:4PhaseDiagram_new} and Fig.~\ref{fig:phase_coexist} both show that the system properties change significantly as $\omega_0^* \approx 1$.
Therefore, we focus on the case $\omega_0^* \approx 1$ for further investigation. 
\begin{figure}[!htbp]
  \centering
  \includegraphics[width=0.9\linewidth]{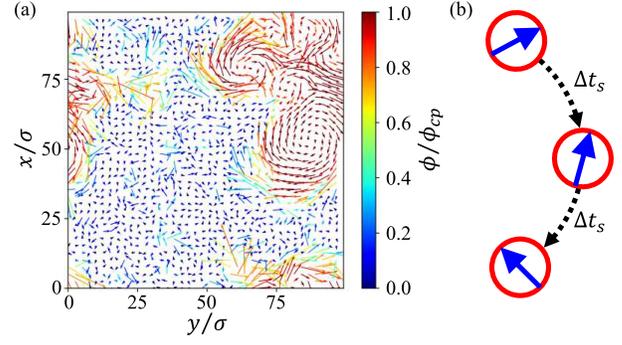}
  \caption[Snapshots for showing clockwise vortices]{(Color online) (a) A snapshot of dimensionless momentum density field at mean area fraction $\Phi = 0.4$, motility $v_0^* = 120$, and angular speed $\omega_0^* = 0.8$. Arrows represent the local momentum density $\boldsymbol{p}$ (see Appendix~\ref{app:field} for details). All arrows are scaled by a factor of $0.5$ for better visual quality. Colors reflect the local area fraction $\phi$ with respect to $\phi_{cp}$. (b) Illustration of the motion of a circle swimmer inside a clockwise vortex. This swimmer moves along a clockwise curve, as indicated by dash black arrows. However, it rotates counterclockwise around its center of mass, as reflected by the blue arrows which represent its orientation.}
  \label{fig:Snapshots}
\end{figure}

\subsection{\label{sec:resultC} Clockwise vortices}
% New Paragraph
%
Intuitively, since the (repulsive) interactions between our circle swimmers are independent of their orientations, each swimmer is expected to move in \textit{counterclockwise} direction as a single swimmer would do (see Sec.~\ref{sec:model}).
However, surprisingly we observe in a certain parameter range ($\Phi = 0.4$, $v_0^* = 120$, and $\omega_0^* = 0.8$) that swimmers inside clusters collectively move along \textit{clockwise} curves, yielding clockwise vortices. 
This is illustrated in Fig.~\ref{fig:Snapshots}(a), as well as in Movie2.avi and Movie3.avi (ESI\dag).
The clockwise vortices are not observed either for the conventional ABPs, or for circle swimmers with high angular speeds, as seen in Movie1.avi for $\omega_0^* = 0$ and Movie4.avi for $\omega_0^* = 10$ (ESI\dag).
We note that, although a swimmer inside a clockwise vortex moves along clockwise curves, it rotates counterclockwise around its center of mass, as sketched in Fig.~\ref{fig:Snapshots}(b).
%

\begin{figure}[!htbp]
    \centering
    \includegraphics[width=0.9\linewidth]{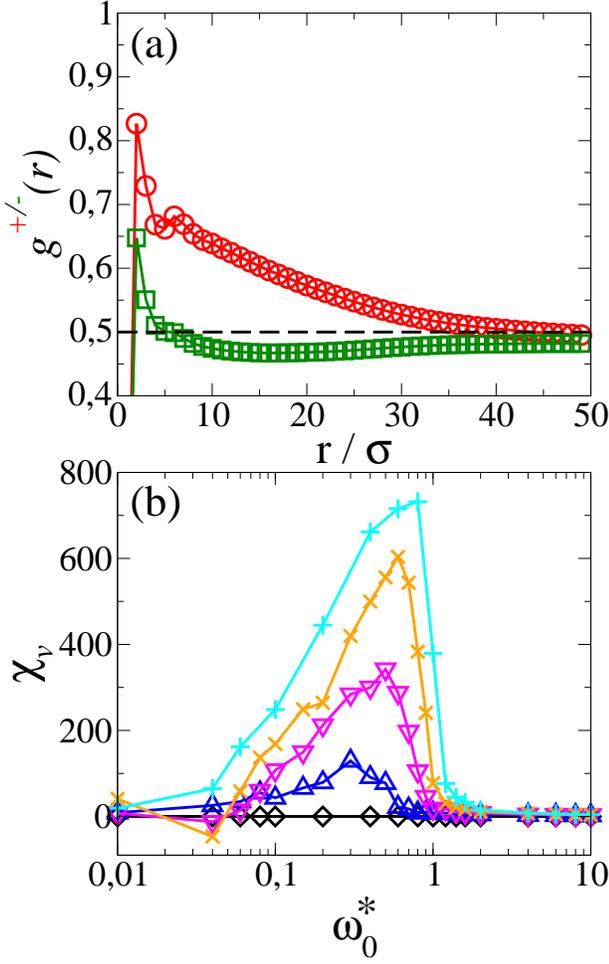}
    \caption[Characterization of clockwise vortices]{(Color online) (a) Pair correlation functions of particles rotating clockwise (denoted as $g^{+}(r)$; red circles) and particles rotating counterclockwise (denoted as $g^{-}(r)$; green squares) at $\Phi = 0.4$, $\omega^*_0 = 0.8$, and $v^*_0 = 90$. (b) Integrated difference between $g^{+}(r)$ and $g^{+}(r)$, $\chi_v$, as a function of angular speed $\omega^{*}_{0}$ for various propulsion speeds $v^*_0 = 0$ (black diamonds), $50$ (blue triangles up), $70$ (magenta triangles down), $90$ (orange crosses), and $110$ (cyan plusses) at $\Phi = 0.4$. The solid lines are guides to the eyes.}
    \label{fig:chi_v}
\end{figure}

% New Paragraph
%
To characterize the clockwise vortices, we calculate two types of pair correlation functions as suggested in ref. \citen{Yeo2015}. 
To this end, we consider the rotation of particle pairs: Imagine a pair of particles $i$, $j$ with relative displacement $\boldsymbol{r}_{ij} = \boldsymbol{r}_{j} - \boldsymbol{r}_{i}$ and velocities $\boldsymbol{v}_i$ and $\boldsymbol{v}_j$ in the $xy$-plane of the coordinate system. 
The relative velocity is thus defined as $\boldsymbol{v}_{ij} = \boldsymbol{v}_j - \boldsymbol{v}_i$. 
We can then consider the quantity
\begin{equation}
\label{eqn:rotate_sign}
R(\boldsymbol{r}_{ij}, \boldsymbol{v}_i, \boldsymbol{v}_j) = \begin{cases}
1\text{,} &\text{if $\left(\boldsymbol{v}_{ij}\times\boldsymbol{r}_{ij}\right)\cdot\widehat{\boldsymbol{z}} > 0$}\\
-1\text{,} &\text{else.}
\end{cases}
\end{equation}
Pairs of particles with clockwise rotation are characterized by $R = 1$, while the counterclockwise rotating pairs are characterized by $R = -1$.
The pair correlation functions related to clockwise (+) and counterclockwise (--) rotation are defined by
\begin{equation}
\label{eqn:num_density}
g^{\pm}(r) = 
\left\langle 
  \frac{1}{N}
    \sum\limits_{i=1}^N 
    \sum \limits_{j=1,j \neq i}^N
      \frac{\vert R \pm 1 \vert}{2}
      \frac{ \delta 
        \left(
          r - \vert \boldsymbol{r}_{ij} \vert
        \right)
      }{ 
        2 \pi r \rho
      } 
\right\rangle\text{,}
\end{equation}
where $\delta(x)$ is the Dirac function and $\rho = N/L^2$ is the number density. 
Examples for $g^{\pm}\left(r\right)$ in the relevant parameter range are given in Fig.~\ref{fig:chi_v}(a).
For a broad range of distance ($r / \sigma \lesssim 30$), there are more pairs rotating clockwise than counterclockwise.
At even larger distances, $g^{+}(r)$ and $g^{-}(r)$ gradually converge to the same value of $0.5$, suggesting that the direction of rotation becomes uncorrelated at large distances. 
As an overall measure of the magnitude of their correlations, we consider the integrated quantity 
\begin{equation}
  \chi_{v} = \rho \int_{0}^{L / 2} \Big[ g^{+}\left( r \right) - g^{-}\left( r \right) \Big]2 \pi r \text{d} r.
\end{equation}
Figure~\ref{fig:chi_v}(b) shows that $\chi_{v}$ has a maximum for intermediate angular speeds $\omega_0^{*} \approx 0.5$. 
The presence of this maximum clearly indicates that the dominance of clockwise vortices in a certain parameter regime.
%

\begin{figure}[!htbp]
    \centering
    \includegraphics[width=0.9\linewidth]{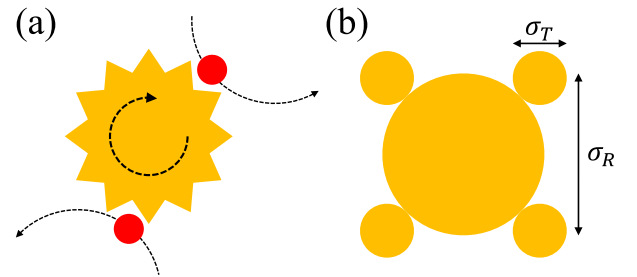}
    \caption[Illustration of the direction of vortex rotation]{(Color online) (a) Illustration of the ``gear'' argument for clockwise rotation of an entire cluster. Individual circle swimmers (red) moving counterclockwise interact with the teeth of the gear (yellow) and thus yield clockwise motion of the gear. The dashed arrows represent the motion of the corresponding objects. (b) A sketch of the passive gear considered in our simulations.}
    \label{fig:gear}
\end{figure}

% New Paragraph
%
The physical origin of this intriguing behavior may be understood by the following simplified picture: 
Given that the shape of a cluster is roughly circular with some defects, we can view this cluster as a ``gear'', as illustrated in Fig.~\ref{fig:gear}(a).
The motion of the gear is perturbed by thermal fluctuations and, more importantly, by steric repulsion with surrounding isolated circle swimmers moving counterclockwise.
In particular, when the circle swimmers collide with a tooth of the gear, they exert unequal impulses at both sides of the tooth.
Taken altogether, this introduces clockwise rotation to the gear.
%

%% New Paragraph
%
To support our argument, we have performed Brownian dynamics simulations of a passive model gear immersed in a suspension of circle swimmers. 
The model gear is inspired by the ``spinners'' considered in ref. \citen{Nguyen2014}, and it is sketched in Fig.~\ref{fig:gear}(b).
It is composed of 4 disk-shaped teeth (T) and one central root (R), where the corresponding diameters are $\sigma_T = 2 \sigma$ and $\sigma_R = 6 \sigma$. 
The gear and swimmers interact merely via steric repulsion. 
Further simulation details are given in Appendix~\ref{app:gear}.
%

\begin{figure}[!htbp]
  \centering
  \includegraphics[width=0.9\linewidth]{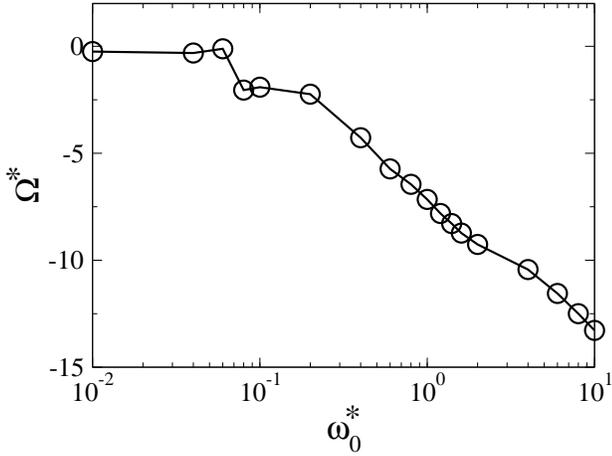}
  \vspace{-5pt}
  \caption{Angular velocity $\Omega^*$ of the passive gear as a function of angular speed $\omega_0^*$ at mean area fraction $\Phi = 0.2$ and motility $v_0^* = 120$.}
  \label{fig:Omega}
\end{figure}

%% New Paragraph
%
To investigate whether the individual swimmers indeed induce a rotation, we plot in Fig.~\ref{fig:Omega} the dimensionless angular speed $\Omega^* = \Omega / D_r$ of the gear as a function of propulsion angular speed $\omega_0^*$ at mean area fraction $\Phi = 0.2$ and motility $v_0^* = 120$. 
At small angular speed of the swimmers ($0 \leq \omega_0^* \lesssim 0.1$), the gear on average does not rotate. 
However, for $\omega_0^* \gtrsim 0.1$ we observe indeed non-zero rotation of the gear in clockwise direction, as indicated by the negative values of $\Omega^*$.
Specifically, the magnitude of $\Omega^*$ grows approximately linearly with $\text{log}\left(\omega_0^*\right)$.
A video example at $\omega_0^* = 1$ is shown in Movie5.avi (ESI\dag).
We conclude that our simplified picture, in which the cluster is considered as a passive gear surrounded by individual circle swimmers, provides indeed a mechanism for the emergence of clockwise vortices.
%

%% New Paragraph
%
Still, one question remaining is whether the clockwise vortices are an artefact of the simulations.
To this end we performed two types of test calculations.
First, to see if the clockwise vortices occur due to finite-sized effects, we performed simulations with different number of particles $N = 2000$ and $N = 10000$ (instead of $N=5000$) at $\Phi = 0.4$, $v_0^* = 120$, and $\omega_0^* = 0.8$.
As it turns out, clockwise vortices still appear either for the smaller or larger $N$.
Second, we tested whether the clockwise vortices are sensitive to the initial conditions.
This is not the case: 
Clockwise vortices still occur, no matter whether we place swimmers on a quadratic lattice or randomly put particles inside the simulation box as an initial configuration.

\section{\label{sec:conclusions} Conclusions}
In the present simulation study, we have investigated the influence of active rotation on the clustering behavior and phase separation of circle swimmers in a two-dimensional geometry.

% New Paragraph
At low mean area fraction ($\Phi = 0.2$), we observed the emergence of small clusters, whose size grows linearly with swimmer motility $v_0^*$. 
As the density increases the clusters transform into giant clusters coexisting with individual circle swimmers.
However, this behavior is observed only at small angular speeds $\omega_0^*$, \textit{i.e.} in the regime where the rotational thermal fluctuations dominate active rotation.
At larger values of $\omega_0^*$, a swimmer can alter its orientation and thus, escape from a cluster, much more rapidly than merely by the rotational diffusion.
This opposes the formation of giant clusters, since number of particles leaving the cluster per unit time increases.
Therefore, we observe a drastic decrease of the size of the largest cluster in the range $\omega_0^* \geq 1$.
%

%% New Paragraph
%
By employing a Voronoi tessellation we have obtained a histogram of the position-resolved, time-averaged local area fractions, which enable us to quantitatively determine motility-induced phase separation and the corresponding binodal curves in the ($v_0^*$, $\phi$) plane.
At fixed mean area fraction $\Phi = 0.4$, we find that active rotation in the range $0 < \omega_0^* \leq 1.2$ generally suppresses motility-induced phase separation by shifting the region of phase coexistence to larger values of $v_0^*$.
Further increase of active rotation ($\omega_0^* > 1.2$) causes the phase separation to disappear.
%

% New Paragraph
%
Moreover, we discovered a novel state characterized by the formation of clockwise vortices at intermediate angular speeds ($\omega_0^* \approx 1$).
We have shown that the underlying mechanism can be captured by a simple argument, where the cluster is considered as a passive ``gear'' surrounded by isolated circle swimmers colliding with the ``gear''.
We should note, however, that this argument does not take into account hydrodynamic interactions (which are neglected in our simulations as well).
For conventional ABPs (no active rotation), it has been reported that hydrodynamic interactions tend to suppress motility-induced clustering. \mbox{\cite{Matas-Navarro2014, Matas-Navarro2015}}
Indeed, compared to the case of conventional ABPs, binodals for hydrodynamically interacting squimers suggest that the motility-induced phase separation occurs at a higher value of P\'eclet number, as shown in Fig. 10 of ref. \citen{Blaschke2016}. 
This indicates the hydrodynamic interactions hinder motility-induced clustering.
%

%% New Paragraph
%
In systems of rotating particles, such hydrodynamic interactions can indeed induce cluster rotation due to the coupling of translational and rotational motion. \cite{Jager2013}
The rotating direction of a cluster is the same as that of a single particle.
However, the vortices observed in the present work rotates oppositely to the rotating direction of a single swimmer, as illustrated by the ``gear'' argument.
Therefore, there is a competition between the ``gear'' mechanism and the impact of translation-rotation coupling, and this presumably opposes the clockwise vortices.
Nevertheless, the detailed impact of hydrodynamic interactions on the collective behavior of circle swimmers remains to be explored by further studies.
%

%% New Paragraph
%
To realize our simplistic model in experiments, one could think of active colloids subject to a rotating magnetic field \cite{Han2017} or chiral Janus particles driven by induced charge electrophoresis. \cite{Mano2017}
Since related experiments often involve electromagnetic interactions, another question concerns the impact of additional (conservative) orientational interactions, an important example being dipolar interactions. 
Investigations in this direction are under way.

\section*{Conflicts of interest}

There are no conflicts to declare.

\section{Appendix}

\subsection{\label{app:steady_state} Details on the cluster analysis}
\begin{figure}[!htbp]
  \centering
  \includegraphics[width=0.9\linewidth]{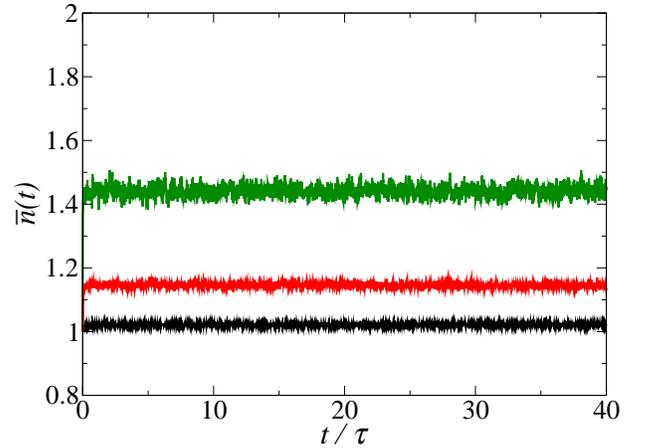}
  \caption{(Color online) Mean cluster size $\bar{n}(t)$ as a function of time at $\Phi = 0.2$ and $\omega_0^* = 0$ for $v_0^* = 0$ (black) , $v_0^* = 30$ (red) and $v_0^* = 120$ (green). Simulations start at $t=0$.}
  \label{fig:meanN_vs_t}
\end{figure}
\begin{figure}[!htbp]
  \centering
  \includegraphics[width=0.9\linewidth]{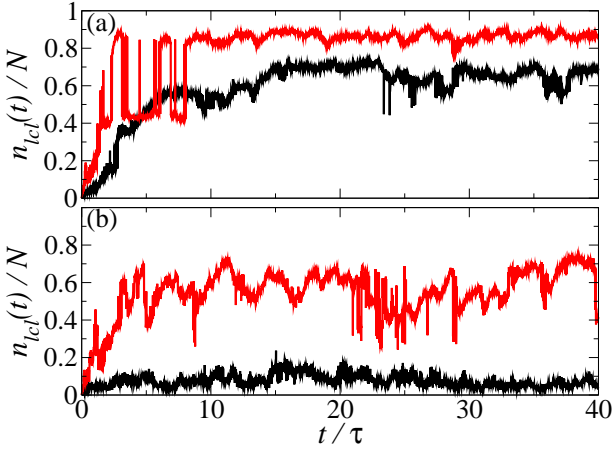}
  \caption{(Color online) Relative size of the largest cluster as a function of time at $\Phi = 0.4$ for $\omega_0^* = 0$ (a) and $\omega_0^* = 1$ (b), where black lines stand for $v_0^* = 60$ (black) and red lines represent $v_0^* = 120$. Simulations start at $t=0$.}
  \label{fig:nlcl_vs_t}
\end{figure}

We define clusters according to a distance criteriion as described in Sec.~\ref{sec:resultA}.
Here we describe some details of the averaging procedure underlying the data in Fig. \ref{fig:meanClust} and \ref{fig:clustr_w0}.
Assuming that there are $N_c(t)$ clusters in the simulation box at a given time $t$, the instantaneous mean cluster size is defined as
\begin{equation}
\bar{n}(t) = \dfrac{\sum_{i=1}^{N_c(t)}n_i(t)}{N_c(t)} \text{,}
\end{equation}
where $n_i(t)$ represents the size of $i$th cluster.
As can be seen in Fig. \ref{fig:meanN_vs_t}, the instantaneous mean cluster size $\bar{n}(t)$ reaches a plateau quickly after the simulation starts, either for passive particles ($v_0^* = 0$), where $\bar{n}(t)$ fluctuates around $1$ (as expected for free particles), swimmer with low motility ($v_0^* = 30$), and for highly-motile swimmers ($v_0^* = 120$).
To ensure that the simulations have approached a steady state, the time average of the mean cluster size $\langle \bar{n} \rangle$ shown in Fig. \ref{fig:meanClust} is calculated for the time interval $20 < t / \tau \leq 40$, \textit{i.e.} the second half of a single simulation.
%

%% New Paragraph
%
Likewise, the size of the instantaneous largest cluster, $n_{lcl}(t)$, is the largest number among $n_i(t)$ for $1 \leq i \leq N_c(t)$.
Figure \ref{fig:nlcl_vs_t} shows $n_{lcl}(t) / N$ as a function of time.
The sudden increase or decrease of $n_{lcl}(t) / N$ at $v_0^* = 120$ can be attributed to a merging of two giant clusters into one, or a breaking of the largest cluster into two clusters.
The size of the instantaneous largest clusters $n_{lcl}(t)$ remain rather stable at $t / \tau > 20$, which indicates that the simulations have reached a steady state.
Therefore, we calculate the time average, $\langle n_{lcl} \rangle/ N$, in Fig. \ref{fig:clustr_w0} for the time interval $20 < t / \tau \leq 40$.

\subsection{\label{app:interface} Identification of the interface between dense and dilute regions}
To identify the interface separating coexisting states, we first define the number of neighbors of a grid point located at $\left(x,y\right)$,
\begin{equation}
\alpha\left(x,y\right) = \Delta L^2 \sum_{j,\; k=-1}^{\:\:\;\;1\;\;\prime}\bar{\phi}\left(x + j \Delta L, y + k \Delta L\right)\text{,}
\end{equation}
where the prime attached to the summation sign indicates that the term $j=k=0$ is omitted.
The interface is then regarded as the set of all grid points which satisfies $\mid\alpha\left(x,y\right) - \alpha_{thres}\mid \leq \delta \alpha_{thres}$.
The threshold $\alpha_{thres} = 4$ and its error $\delta \alpha_{thres} = 0.4$ are chosen such that all these interfacial grid points represent as accurately as possible the interface seen in simulation snapshots (see the insets of Fig.~\ref{fig:interface}). 
An example for the local area fraction as a function of the relative position $d$ to the interface between the dense and dilute region is presented in Fig.~\ref{fig:interface}. 
Here, $\mid d \mid$ is the shortest distance from a grid point to the interface. 
The sign of $d$ is assigned to be positive if $\alpha\left(x,y\right) > \alpha_{thres} + \delta \alpha_{thres}$ (dense region), and negative if $\alpha\left(x,y\right) < \alpha_{thres} - \delta \alpha_{thres}$ (dilute region).

\subsection{\label{app:field} Momentum density field}
\begin{figure}[!htbp]
  \centering
  \includegraphics[width=0.7\linewidth]{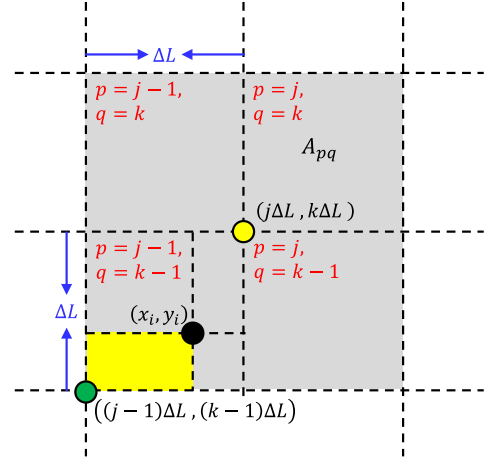}
  \caption{(Color online) Illustration of our mapping procedure of the momentum density $\phi_i\boldsymbol{v}_i$ onto a grid. The green spot represents the location $\left(\left(2p-j+1\right)\Delta L, \left(2q-k+1\right)\Delta L\right)$ for $p = j-1, q = k-1$.}
  \label{fig:interpolation}
\end{figure}

To visualize the rotation of clusters, it is helpful to map the momentum density $\phi_{i}\boldsymbol{v}_{i}$ for the $i$th particle onto a two dimensional grid with mesh size $\Delta L = L/\text{floor}(L/2.5) \approx 2.5\sigma$. 
Each grid vertex takes into account the weighted sum of momentum densities over all swimmers located inside the adjacent mesh cells $A_{pq}$. The momentum density located at grid vertex $\left( j \Delta L, k\Delta L \right)$ is given by
\begin{equation}
\label{eqn:reverse_interplation}
\boldsymbol{p} \left( j \Delta L, k\Delta L \right) = 
  \dfrac{\sigma^2}{\Delta L^2}\sum_{p = j-1}^{j}\sum_{q = k-1}^{k}\sum_{i \in A_{pq}}
  w_i \phi_i \boldsymbol{v}_{i}^* \text{,}
\end{equation}
where the weight
\begin{equation}
  w_i = \vert x_i/\Delta L - \left(2p-j+1\right) \vert
        \vert y_i/\Delta L - \left(2q-k+1\right) \vert
\end{equation}
is essentially the yellow area divided by the area of a mesh cell, see Fig.~\ref{fig:interpolation}.
In eqn~\eqref{eqn:reverse_interplation}, the velocity is given by $\boldsymbol{v}_{i}^* = \left( \Delta \boldsymbol{r}_{i} / \sigma \right)/ \left(\Delta t_s / \tau \right)$ with the time difference $\Delta t_s = 10^{-2} \tau$ and the corresponding displacement $\Delta \boldsymbol{r}_i$.

\subsection{\label{app:gear} Simulation details of a passive gear and circle swimmers}

We consider a passive gear immersed in a suspension of $1000$ circle swimmers. 
As seen in Fig.~\ref{fig:gear}(b), the gear comprises 4 disk-shaped ``teeth'' (T) with diameter $\sigma_T = 2 \sigma$ and one central root (R) with diameter $\sigma_R = 6 \sigma$. All teeth are evenly placed on the surface of the central root. The relative position of each tooth to the central root is fixed to the center-to-center distance $\left(\sigma_T + \sigma_R\right)/2$. The $i$th swimmer and the $j$th component of the gear interact via the Weeks-Chandler-Andersen (WCA) potential. \cite{Weeks1971}
\begin{equation} \label{eqn:wcaC}
U_{WCA}(r_{ij})= 
  \begin{cases}
    4\epsilon \left[
      \left(\dfrac{\sigma_{ij}}{r_{ij}}\right)^{12} - 
      \left(\dfrac{\sigma_{ij}}{r_{ij}}\right)^{6} + 
      \dfrac{1}{4}
    \right]
    \text{,} &\text{if $r_{ij} < r_{c}$,}\\
    0\text{,} &\text{else.}
  \end{cases}
\end{equation}
The cut-off radius is given by $r_{c} = 2^{1/6} \sigma_{ij}$. If the $j$th component is a tooth, we assign the interaction range $\sigma_{ij} = \left( \sigma + \sigma_T \right)/2$. Otherwise, the range is defined as $\sigma_{ij} = \left( \sigma + \sigma_R \right)/2$. We set the interaction strength $\epsilon^{*} = \beta \epsilon = 100$.
The corresponding Brownian dynamics simulations are performed with the translational diffusion coefficient of the gear (g) $D_{t,g} = \sigma D_t / \left(\sigma_T + \sigma_R\right)$, and the rotational counterpart $D_{r,g} = D_{t,g}/ \left(\sigma_T + \sigma_R\right)^2$. 

\subsection{\label{app:movie} Movie descriptions}

The details of each movie are provided below:

\textbf{Movie1.avi}: Brownian dynamics simulations of $5000$ circle swimmers at $\Phi = 0.4$, $v_0^* = 120$, and $\omega_0^* = 0.0$. The color of each swimmer is randomly assigned for monitoring its motion. One second in the video corresponds to $0.25 \tau$ in simulations.

\textbf{Movie2.avi}: Brownian dynamics simulations of $5000$ circle swimmers at $\Phi = 0.4$, $v_0^* = 120$, and $\omega_0^* = 0.8$. The color of each swimmer is randomly assigned for monitoring its motion. One second in the video corresponds to $0.25 \tau$ in simulations.

\textbf{Movie3.avi}: Momentum density field of a $39 \times 39$ grid mapped from Brownian dynamics simulation of $5000$ circle swimmers at $\Phi = 0.4$, $v_0^* = 120$, and $\omega_0^* = 0.8$. The color of each arrow is chosen according to the value of the local area fraction divided by effective close-packing area fraction. Each arrow represents the local momentum density with the arrow length scaled by a factor of $0.5$ for better visibility. One second in the video corresponds to $0.25 \tau$ in simulations.

\textbf{Movie4.avi}: Brownian dynamics simulations of $5000$ circle swimmers at $\Phi = 0.4$, $v_0^* = 120$, and $\omega_0^* = 10$. The color of each swimmer is randomly assigned for monitoring its motion. One second in the video corresponds to $0.25 \tau$ in simulations.

\textbf{Movie5.avi}: Brownian dynamics simulations of a single passive gear and $1000$ circle swimmers at $\Phi = 0.2$, $v_0^* = 120$, and $\omega_0^* = 1$. One second in the video corresponds to $0.25 \tau$ in simulations.

\section*{Acknowledgements}

The authors would like to thank Deutsche Forschungsgemeinschaft for the financial support from GRK 1524 (DFG No. 599982). We also thank H. Stark for fruitful discussions.

%%%END OF MAIN TEXT%%%


\balance


%%%REFERENCES%%%
\bibliography{circle_swimmers} %You need to replace "rsc" on this line with the name of your .bib file
\bibliographystyle{rsc} %the RSC's .bst file

\end{document}